\documentclass[journal]{IEEEtran}
\usepackage{amsmath,amsfonts}
\usepackage{algorithmic}
\usepackage{array}
\usepackage{textcomp}
\usepackage{stfloats}
\usepackage{url}
\usepackage{verbatim}
\usepackage{graphicx}
\def\BibTeX{{\rm B\kern-.05em{\sc i\kern-.025em b}\kern-.08em
    T\kern-.1667em\lower.7ex\hbox{E}\kern-.125emX}}
\usepackage{balance}
\usepackage{makecell}
\usepackage[square, comma, sort&compress, numbers]{natbib}
\usepackage{mathrsfs}
\usepackage{booktabs}
\usepackage{xcolor}
\usepackage{subfigure}

\ifCLASSINFOpdf
\else
\fi

\begin{document}
\title{Field of View Expansion for Resonant Beam Information and Power Transfer}
\author{Shun Han,
Wen Fang,
Mingqing Liu,
Mengyuan Xu,
Shuaifan Xia,
and Qingwen Liu,~\IEEEmembership{Senior Member,~IEEE}
\thanks{S.~Han, M.~Liu, M. Xu, S. Xia, and Q.~Liu are with the College of Electronic and Information Engineering, Tongji University, Shanghai 201804, China (e-mail: hanshun@tongji.edu.cn, clare@tongji.edu.cn, xumy@tongji.edu.cn, collinxia@tongji.edu.cn, and qliu@tongji.edu.cn).}
\thanks{W. Fang is with the School of Electronic Information and Electrical Engineering, Shanghai Jiao Tong University, Shanghai 200240, China (e-mail: wendyfang@sjtu.edu.cn).}}

%\markboth{Journal of \LaTeX\ Class Files,~Vol.~18, No.~9, September~2020}%
%{Resonant beam simultaneous wireless %information and power
%transfer based on FOV improved %retro-reflector}

\maketitle

\begin{abstract}
Simultaneous wireless information and power transfer (SWIPT) leverages lightwave as the wireless transmission medium, emerging as a promising technology in future Internet of Things (IoT) scenarios. 
The use of retro-reflectors in constructing spatially separated laser resonators (SSLR) enables a self-aligning wireless transmission system with the self-reproducing resonant beam, i.e. resonant beam system (RBS).
However, its effective Field of View (FoV) is physically limited by the size of retro-reflectors and still requires significant improvement. 
This restricts the transmitter from providing seamless wireless connectivity and power supply to receivers within a large dynamic movement range. 
In this paper, we propose an FoV-enlarged resonant beam system operating at a distance of meters by incorporating a telescope.
The telescope plays a crucial role in minimizing the extra loss inflicted on the gain medium, which typically arises from the deviation of the resonant beam within the cavity. 
Further, we construct the proposed telescope-based RBS and experimentally demonstrate that the design could expand the FoV to 28$^\circ$ over 1~m transmission distance, which is about triple that of the ordinary RBS design.

\end{abstract}


\section{Introduction}
\IEEEPARstart{I}{n} the era of future 6G, AR/VR technologies, robotics, and IoT technologies will play key roles in improving people's working efficiency, quality of life, and living standards~\cite{koonen_beam-steered_2021, 6951347}. 
Beyond the reliance on communication infrastructure, sustained battery life to ensure uninterrupted operation of the device throughout the day is also critical.
Simultaneous wireless information and power transfer (SWIPT) technology, which employs radio frequency (RF) as the carrier, facilitates wireless communication with mobile devices while ensuring continuous power transfer. 
This approach significantly enhances the endurance of mobile communication devices by providing a sustained energy supply.
Nevertheless, the development of SWIPT is constrained by problems with RF spectrum scarcity and low RF energy harvesting efficiency. 

\begin{figure}[!t]
\centering
\includegraphics[scale=0.9]{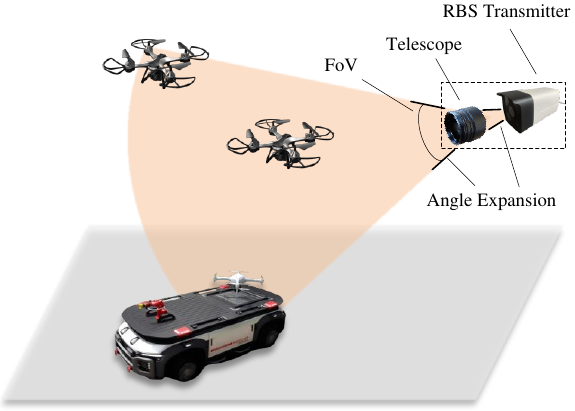}
\caption{The example of application scenario for resonant beam SLIPT system}
\label{scenario}
\end{figure}

The simultaneous lightwave information and power transfer (SLIPT) technology, proposed in recent years, can simultaneously achieve high-bandwidth signal transmission and high-power energy transfer, meeting the needs of mobile devices and applications for network and battery life~\cite{8322210}.
In SLIPT systems, light-emitting diodes (LEDs) and laser sources are generally used as carriers for energy and signal transmission~\cite{palitharathna_lightwave_2022, pan_simultaneous_2019, ma_simultaneous_2019}. 
Using lightwave as a transmission carrier for communications can circumvent electromagnetic interference and possess a broader communication bandwidth. 
On the subject of power transfer, it has advanced transmitter devices and energy harvesting equipment, enabling higher end-to-end efficiency.
A study in~\cite{shi_si-substrate_2022} has validated a 1~m light communication link that adopts a self-designed $4\times4$ $50~ \mu {\rm{m}}$ $\mu$PD and a 450~nm laser at a rate of 10.14~Gbps.
Another work in~\cite{fakidis_indoor_2016} proposed a laser-based wireless power transfer system with 42 lasers embedded in the transmitter, which transmits 7.2~W optical power to a receiver 30~m away with a geometric loss of only 2\%.
Despite the advantages of SLIPT technology, such as high system efficiency, powerful transfer capabilities, and rapid communication rates, it still faces challenges in balancing large FoV with high transmission efficiency. 
Utilizing lasers with a narrow divergence angle as the light source can establish transmission links only within a limited FoV. 
Unfortunately, strict alignment of the transmitter and receiver is generally required~\cite{9511521,9127159}.
Employing LEDs as light sources or diffusing lasers with a uniform light device can effectively expand the system's FoV~\cite{alshaibani_wide-field--view_2023}, but this design brings the reduction of the system's transmission efficiency.
A standard solution to such problems involves the adoption of the acquisition, tracking, and pointing (ATP) mechanisms~\cite{wang_wireless_2019, liu_charging_2022}.
This method enables the connection between the transmitter and receiver by utilizing narrow beams through mutual direction acquisition, continuous tracking, and eventual alignment, thus allowing the transceivers to establish a connection within a larger FoV.
In free space optical communications, it is expected to employ an ATP mechanism based on gimbal systems and fast steering mirrors for aligning the transceiver ends, enabling the transmission of narrow beams~\cite{kingsbury2017fast}. 
This approach can achieve a wider FoV, but the size and power consumption of the equipment make it unsuitable for IoT mobile devices.
The ATP mechanism based on liquid crystal technology boasts advantages such as low cost, low power consumption, and light weight~\cite{cryst9060292}. 
However, this method is limited in its ability to deflect the beam angle, making it challenging to achieve a large coverage area~\cite{pouch2006liquid}.
Moreover, for the real-time alignment between the transmitter and a receiver in motion, it is imperative to deploy high-precision control modules at both the transmitting and receiving ends~\cite{urabe_high_2012}.
This implementation escalates the overall computational and hardware expenditure.

Resonant beam transfer system (RBS) is one of the wireless transmission systems featuring automatic alignment capabilities~\cite{8999738,9374095}.
It can transmit narrow-beam lasers within a certain FoV without the necessity for additional control hardware and algorithms~\cite{10134572}. 
The transceivers consist of retro-reflectors, allowing the transmission link to be established as long as the receiver and transmitter are within each other's FoV~\cite{sheng_adaptive_2023}.
However, the pump light and the resonant beam need to coincide in the gain medium at the transmitter. It is challenging to improve the FoV of the system, especially the FoV of the transmitter. 
Currently, in experimental setups, the maximum achievable FoV for the transmitter is only about 6$^\circ$, which is insufficient to meet the demands of applications~\cite{liu_charging_2022}. 
This limitation highlights a significant challenge in the broader implementation of such systems in IoT scenarios.

To address the constraints on allowable incident angles due to the gain medium,
this paper introduces an improved RBS transmitter equipped with an adaptive angle control module, which can adjust the relationship between the incident angle of the resonant beam at the transmitter and the angle at which the resonant beam hits the gain medium. 
This advancement allows us to overcome existing limitations, leading to a significant improvement in the system's FoV.
The contributions of our work are as follows.
\begin{enumerate}{}{}
\item{We redesigned the transmitter of the RBS and conducted experiments to validate the feasibility of the improved RBS with the telescope. 
Utilizing the enhanced RBS transmitter, the system's transmitter FoV can be expanded from 10$^\circ$ to 28$^\circ$.}
\item{Based on the RBS we designed, we established a system stability analysis model and analyzed the maximum stable working distance of the system under various conditions.}
\end{enumerate}

The remainder of this paper is organized as follows. 
Section II delved into analyzing the key factors limiting the FoV enhancement in RBS. Based on this analysis, we proposed a new system design to increase the system's FoV. 
In Section III, We analyze the results of our transmission experiments and the performance comparison results. Afterward, we present a scheme to improve the FoV of the resonant beam SLIPT system in Section IV. Finally, we conclude in Section V.

\section{System Model}

In this section, we dissect the pivotal constraints that curtailed the system's FoV in previous studies.
Subsequently, an advanced enhancement strategy, leveraging the telescope, is proposed to augment the system's FoV.
Finally, we analyze the transmission performance to demonstrate the feasibility of the proposed system.

\subsection{Factors Influencing the FoV of the RBS}

\begin{figure}[!t]
\centering
\includegraphics[scale=0.95]{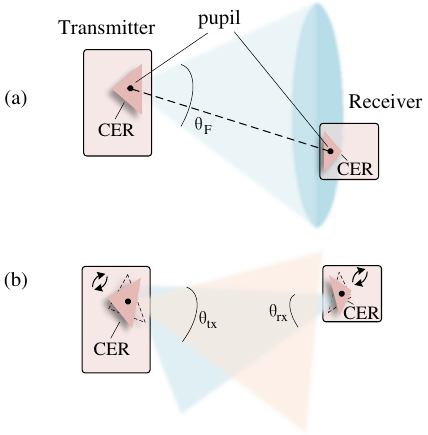}
\caption{Schematic of (a) the resonant beam transfer system’s FoV; (b) the receiver and the transmitter FoV.}
\label{introduction}
\end{figure}

In prior research, the RBS predominantly comprised spatially separated laser resonators (SSLR), which a pair of retro-reflectors could constitute. 
The SSLR is characterized as a misalignment-tolerant cavity~\cite{9410573,8749948}, signifying that stable oscillations and laser output can be sustained even when deviations and rotations are present in the transmitter and receiver. 
Hence, the misalignment tolerance of the SSLR is a vital evaluation criterion of its performance.
In this paper, the misalignment tolerance of the SSLR is expressed by the FoV of the RBS.
As shown in Fig.~\ref{introduction}, in an SSLR constituted by CERs, two optical pupils inherently exist within the cavity.
We define the FoV as the maximum displacement angle of the receiver's pupil relative to that of the transmitter's pupil while ensuring stable oscillations within the cavity.
For instance, the angle $\theta_{\rm{F}}$ in Fig.~\ref{introduction}(a) denotes the magnitude of the system's FoV. At the same time, the blue conical region illustrates the permissible movement range of the receiver within the FoV.

The FoV of the RBS system is influenced by several factors, including the FoV of the CERs at the transmitter and receiver~\cite{sheng_adaptive_2023}, the resonant beam pumping structure, and the intra-cavity loss. 
Among these, the FoV of the CERs at the transmitter and receiver determines the theoretical maximum FoV of the system. 
Therefore, designing CER with as large an FoV as possible constitutes the initial step in the design of RBS systems.  
However, the experimentally measured FoV often falls short of the system's theoretical maximum, making it crucial to investigate further the factors that influence the system's FoV, to progressively align the actual FoV with the theoretical maximum.
Different resonant beam pumping structures offer various pumping modes suitable for different resonator structures, and an appropriate pumping structure can help the system achieve a larger FoV. 
Intra-cavity loss primarily includes the spatial transmission loss of the resonant beam, diffraction loss through various optical components, reflection and absorption loss through lenses, and loss through the gain medium. 
When intra-cavity loss exceeds the pump gain, the resonator will fail to oscillate, rendering the RBS system inoperative. 
Thus, it is imperative to ensure that intra-cavity loss remains at a low level across different FoV angles.
Optimizing the resonator structure can reduce transmission and diffraction losses, though significant improvements are challenging to achieve. 
The losses due to reflection and absorption by lenses are relatively fixed and minimal. 
However, the loss of resonant beam in the gain medium can vary significantly with changes in the SSLR structure, making it a critical component of intra-cavity loss.

\subsubsection{FoV of the CERs at the transceivers}
The design of the CER is one of the factors affecting the FoV of the RBS system.
As shown in Fig.~\ref{introduction}(b), the CER at receiver, can rotate around the optical pupil at a large angle while maintaining the stability of the RBS. 
This angle of rotation $\theta_{\rm{rx}}$, achieved without moving the CER at the transmitter, is referred to as the receiver FoV. 
Similarly, by keeping the CER at the receiver stationary, the CER at the transmitter can rotate around the optical pupil to the maximum angle $\theta_{\rm{tx}}$, known as the transmitter FoV.
Stable connections between the transceivers are achieved only when the effective coverage areas of the transmitter and receiver overlap.
The FoV of the RBS transmitter and receiver are respectively influenced by the FoV of the CERs at the transmitter and receiver.
The CER structure usually consists of a mirror and a lens, as shown in Fig.~\ref{challenges}. 
There will be an optical pupil in front of the lens of the CER, and the light incident from the pupil within the FoV of the CER will be reflected in the opposite direction.
The FoV of the CER is given by
\begin{equation}
    \label{FoV01}
\theta_{\rm{F}} = 2{\rm{arctan}}(l_{\rm{a}}/f),
\end{equation}
where $l_{\rm{a}}$ is the radius of the aperture of the lens in the CER, and $f$ is the focus length of the lens in the CER.

\subsubsection{Resonant beam pumping structure}

\begin{figure}[!t]
    \centering
    \includegraphics[scale=0.9]{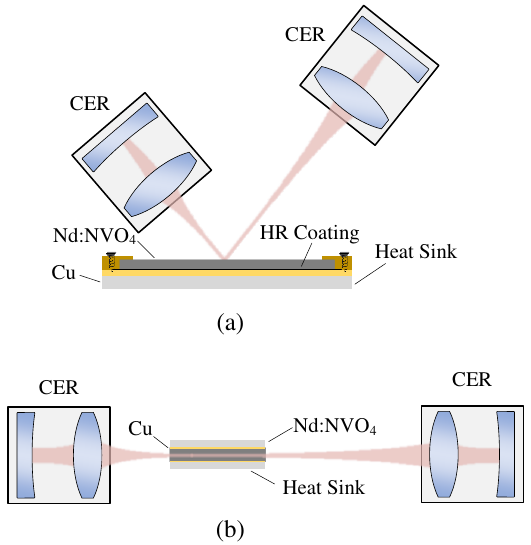}
    \caption{Schematic diagram of the resonant beam transmission system featuring: (a) reflective gain medium or (b) transmissive gain medium. (HR Coating: High-reflectivity Coating)}
    \label{challenges}
\end{figure}

Based on the different forms of gain crystals, the design of RBS can be broadly categorized into two types: rod crystals featured RBS and disk crystals feature RBS, as depicted in Fig.~\ref{challenges}.
In most RBSs supporting receiver mobility~\cite{liu_charging_2022,zhang_wireless_2023,yun_stability_2021,zhang_2m-distance_2022}, the structure commonly adopted is the disk-crystal configuration depicted in Fig.~\ref{challenges}(a).
One distinct advantage of this configuration is the ease with which semiconductor heat sinks can be installed behind the gain medium, facilitating efficient thermal dissipation. 
Furthermore, strategically placing the gain medium at the pupil location ensures consistent overlap between the pump light and the resonant beam within the gain medium, effectively enhancing the system's overlap efficiency.

Furthermore, a segment of the research has employed rod-shaped crystals as the gain medium within the cavity, as is shown in Fig.~\ref{challenges}~(b)~\cite{liu_large-range_2022, SHENG2022108011}. 
Under this scheme, if the pump light is introduced through the end face, any radial movement of the CER at the receiving end can result in a misalignment between the resonant beam and the pump light, subsequently impairing the system's transmission efficiency and constraining the permissible movement angle of the receiver. 
Under this configuration, side-pumping can also be employed~\cite{10288532}. 
This pumping approach necessitates full pumping of the rod-shaped crystal, which can significantly compromise the system's transmission efficiency. 
Regardless of the pumping approach, the system's FoV is significantly constrained.
Consequently, system designs adopting this pumping scheme primarily emphasize enhancing the rotational angle of the CER at the receiver (which is receiver FoV) rather than its radial movement range~\cite{zhang_wireless_2023}.
In summary, the setup utilizing the reflective disk-crystal demonstrates superiority over the rod-crystal setup in FOV performance.

\begin{figure}[!t]
\centering
\includegraphics[scale=0.9]{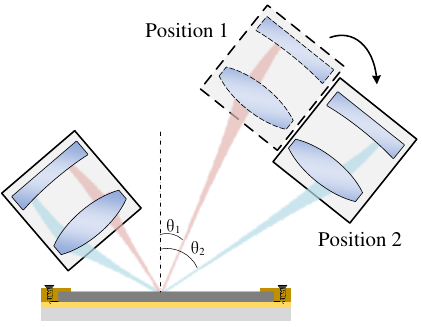}
\caption{Schematic illustration of the change in the incidence angle of the resonant beam passing through the gain medium when the CER at receiver undergoes radial movement relative to the transmitter.}
\label{challenges02}
\end{figure}

\subsubsection{Loss of the resonant beam in the gain medium}

For the disk-crystal configuration, the performance of the HR coating behind the gain medium plays a pivotal role in influencing the system's FoV.
Upon radial displacement of the receiving end relative to the transmitting counterpart, there ensues a continual alteration in the incidence angle of the resonant beam traversing the gain medium as shown in Fig.~\ref{challenges02}.
This causes a change in the reflectivity of the HR coating.
In traditional resonant beam transmitters with a reflective gain medium, the gain medium is positioned at the pupil of the CER.
Suppose the angle between the gain medium and the CER optical axis is $\theta_{\rm{G}}$. 
In that case, the beam entering from the pupil within the FoV of the CER has an angle with the gain medium that falls within the range 
\begin{equation}
    [\theta_{\rm{G}}-{\rm{arctan}}(l_{\rm{a}}/f), \theta_{\rm{G}}+{\rm{arctan}}(l_{\rm{a}}/f)]. 
\end{equation}

HR coatings are typically tailored for a designated angular range and exhibit diminishing reflectivity with increasing angular deviations~\cite{OpticalFilters}.
We have conducted experiments to measure the reflectivity of the gain medium in the variation of the incident angle as in Fig.~\ref{challenges02}, which was coated with a $45^\circ$ HR film.
The experimental results in Fig.~\ref{system01}(d) revealed that HR coatings can only provide a relatively high reflectivity (greater than 90\%) within a narrow angular range (from $31^\circ$ to $52^\circ$), which is smaller than the FoV of standard CERs~\cite{han_transmitter_2023}.
As the incident angle is further expanded, the transmission loss within the SSLR increases accordingly.
Consequentially, this necessitates an elevation in the threshold of the pump light power.
Higher pump light power leads to more severe thermal lens effects, making it difficult for the SSLR to maintain stable oscillations. 
Therefore, the FoV of the transmitter is hard to exceed $20^\circ$.
Given the considerable angular sensitivity of the reflective HR coating, the paramount challenge lies in augmenting the system's FoV within these limitations.

\subsection{System Design}

\begin{figure*}[!t]
\centering
\includegraphics[scale=1.2]{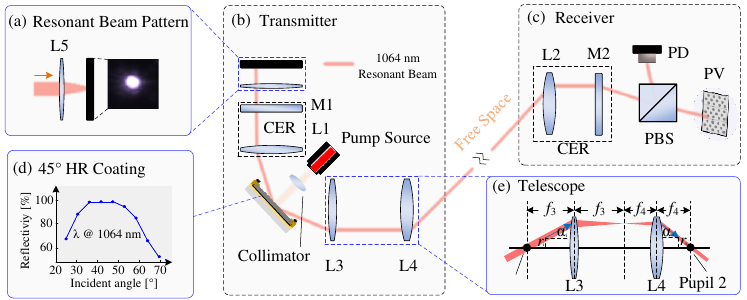}
\caption{Schematic of the resonant beam transfer system (PBS: Polarizing beam splitter; PD: photodetector; PV: photovoltaic). Specifically, (a) resonant beam pattern on the CMOS, (b) depiction of the transmitter, (c) depiction of the receiver, (d) reflectivity of the $45^\circ$ HR coating, and (e) depiction of the telescope.}
\label{system01}
\end{figure*}

In this paper, we introduce the angle amplified resonant beam transmitter (AA-RBT), which effectively mitigates the limitations imposed on the system's FoV by the reflectivity of the gain medium.
The overall system configuration is illustrated in Fig.~\ref{system01}.
The transmitter of the proposed system, serving as a base station, is typically positioned in a fixed location that is less susceptible to obstructions. 
The receiver is a compact mobile terminal conveniently mounted on portable devices.
The transmitter encompasses a CER constituted by the lens (L1) and mirror (M1), a pump source, a telescope structure formed by the lens (L3) and lens (L4), and a gain medium.
The receiving end consists of another CER, a polarizing beam splitter (PBS), a photodetector (PD), and a photovoltaic (PV).

In the SSLR, the resonant beam inside the cavity is first amplified by passing through the gain medium fixed at the pupil of the transmitter CER. 
It then undergoes angular magnification through the telescope structure, and after traversing through free space, it finally reaches the receiver CER. 
The receiver CER reflects the resonant beam to the transmitter. 
At this point, the resonant beam oscillates continuously within the resonant cavity, experiencing amplification and in-cavity losses with each round trip. 
Once the oscillations in the resonant cavity become stable, the receiver CER outputs the steady resonant beam into the PBS. 
The PBS separates a portion of the resonant beam used for communication and directs it into the PD, while the remaining beam, used for power transmission, is directed into the PD for energy harvest.
Due to the SSLR comprising CERs, the system possesses self-aligning characteristics. 
Consequently, when the receiver CER moves freely within the system's FoV, it does not affect the stable operation of the resonant beam transfer system.

\subsection{FoV Enhancement with the Telescope Structure}

In the following, we will analyze the impact of the telescope structure on the system's FOV using the ABCD analysis method.
We depict the telescope structure of the proposed design as in Fig.~\ref{system01}(e).
The plane located at a distance $f_3$ in front of lens L3 is designated as the initial plane of the telescope structure, and the plane situated at a distance $f_4$ behind lens L4 is defined as the termination plane of the structure.
The symbols $f_3$ and $f_4$ denote the focal length of lens L3 and L4, respectively. 
We use a tuple $[r_i \alpha_i]^{\rm{T}}$ to express the beam incident on the initial plane. 
After traversing the telescope structure that is mathematically described by a transfer matrix ${\rm{M_T}}$, the beam can be expressed as  

\begin{equation}
\label{1}
\left[ \begin{array}{l}{r_{\rm{o}}}\\\alpha_{\rm{o}}\end{array} \right] = {{\rm{M_T}}}\left[ \begin{array}{l}{r_{\rm{i}}}\\\alpha_{\rm{i}}\end{array} \right],
\end{equation}
where $r_{\rm{o}}$ represents the ordinate at the point of intersection between the incident ray and the incident plane, while $r_{\rm{i}}$ signifies the ordinate at the intersection point between the emergent rays and the exit plane. 
On the other hand, $\alpha_{\rm{o}}$ and $\alpha_{\rm{i}}$ denote the angles of the incident ray and emergent rays, respectively. The matrix $\rm{M_T}$ is the transfer matrix of the telescope structure and can be expressed as

\begin{align}
{{\rm{M}}_{\rm{T}}} = & \left[ {\begin{array}{*{20}{c}}
1&{{f_4}}\\
0&1
\end{array}} \right]\left[ {\begin{array}{*{20}{c}}
1&0\\
{ - \frac{1}{{{f_4}}}}&1
\end{array}} \right]  \left[ {\begin{array}{*{20}{c}}
1&{{f_3} + {f_4}}\\
0&1
\end{array}} \right]\nonumber \\
& \times\left[ {\begin{array}{*{20}{c}}
1&0\\
{ - \frac{1}{{{f_3}}}}&1
\end{array}} \right] \left[ {\begin{array}{*{20}{c}}
1&{{f_3}}\\
0&1
\end{array}} \right] \nonumber\\
= &\left[ {\begin{array}{*{20}{c}}
-\frac{f_4}{f
_3} & 0 \\0 & -\frac{f_3}{f_4}
\end{array}} \right]=
\left[ {\begin{array}{*{20}{c}}
-\frac{1}{n_2} & 0 \\0 & -n_2
\end{array}} \right],
\end{align}
where $n_2 = \frac{f_3}{f_4}$.
When the incident light enters from pupil 1 of the telescope structure, the transfer segment through the entire telescope can be represented as

\begin{equation}
\begin{aligned}
\left[ \begin{array}{l}{r_{\rm{o}}}\\\alpha_{\rm{o}}\end{array} \right] 
= & \left[ {\begin{array}{*{20}{c}}
-\frac{1}{n_2} & 0 \\0 & -n_2
\end{array}} \right]\left[ \begin{array}{l}{0}\\\alpha_{\rm{i}}\end{array} \right]\\
=&\left[ {\begin{array}{*{20}{c}}
0\\
{ -n_2{\alpha _i}}
\end{array}} \right].
\end{aligned}
\end{equation}
This implies that the emergent light consistently traverses the pupil 2 of the telescope structure.
Moreover, considering $\left| n_2 \right| >1$, the angular deviation of the outgoing light relative to the optical axis surpasses that of the incident light for the same axis.
Therefore, the telescope structure can amplify the angle of the incident beam, and the magnification factor is equal to $n_2$.

\begin{figure*}[!t]
    \centering
    \includegraphics[scale=0.91]{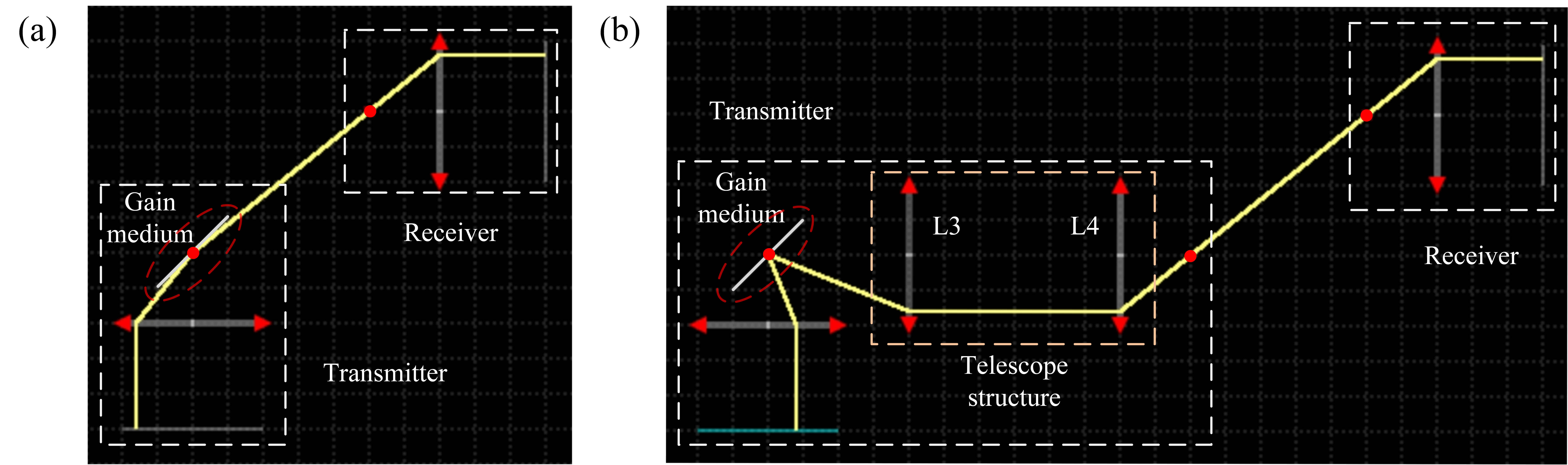}
    \caption{Simulation diagram of SSLR using ray-tracing optics simulator Ray-Optics (a) without telescope structure; (b) with telescope structure.}
    \label{system03}
\end{figure*}

When beam entering from the pupil 2 of the AA-RBT within the angular range $[-{\rm{arctan}}(l_{\rm{a}}/f),{\rm{arctan}}(l_{\rm{a}}/f)]$, impinges upon the gain medium, the angle between the beam and the gain medium fall within the range 
\begin{equation}
    \left[\theta_{\rm{G}}-\frac{{\rm{arctan}}(l_{\rm{a}}/f)}{\left|n_2\right|}, \theta_{\rm{G}}+\frac{{\rm{arctan}}(l_{\rm{a}}/f)}{\left|n_2\right|}\right]. 
\end{equation}
It can be seen from the change in the incident angle range of the resonant beam on the gain medium that the angle amplification of the telescope structure compresses the change range of the incident angle.
Since a telescope structure is added to the transmitter, the FoV of the transmitter will also change.
The theoretical maximum FoV of AA-RBT $\theta_{\rm{A}}$ is given by
\begin{equation}
\label{theta}
\theta_{\rm{A}} = {\rm{min}}\left[2{\rm{arctan}}(\frac{l_{t}}{f_4}), 2{\rm{arctan}}(\frac{{{\left| n_2 \right|l_{\rm{a}}}}}{{{f_1}}})\right],
\end{equation}
where $l_{\rm{t}}$ is the radius of aperture of the telescope structure.
Using the ray-tracing optics simulator Ray-Optics, Fig.~\ref{system03} showcases the transmission of the resonant beam within the SSLR.
In this simulation, the lenses in the resonant cavity are treated as ideal thin lenses.
By incorporating the telescope structure, the incidence angle of the resonant beam on the gain medium is reduced by a factor of $\left| n_2\right|$ under the same transmission distance and radial displacement.

\begin{figure*}[!b]
\hrulefill
\begin{equation}
\label{matrix1}
\begin{aligned}
{{\rm{M}}_s} &= \left[ {\begin{array}{*{20}{c}}
1&{{l_1}}\\
0&1
\end{array}} \right]\left[ {\begin{array}{*{20}{c}}
1&0\\
-\frac{1}{f_1}&1
\end{array}} \right]\left[ {\begin{array}{*{20}{c}}
1&{{f_1} + {f_3}}\\
0&1
\end{array}} \right]\left[ {\begin{array}{*{20}{c}}
1&0\\
-\frac{1}{f_3}&1
\end{array}} \right] \left[ {\begin{array}{*{20}{c}}
1&{{f_3} + {f_4}}\\
0&1
\end{array}} \right]\left[ {\begin{array}{*{20}{c}}
1&0\\
-\frac{1}{f_4}&1
\end{array}} \right]\left[ {\begin{array}{*{20}{c}}
1&{{f_4}}\\
0&1
\end{array}} \right]\left[ {\begin{array}{*{20}{c}}
1&d\\
0&1
\end{array}} \right]\\
 &\times \left[ {\begin{array}{*{20}{c}}
1&{{f_2}}\\
0&1
\end{array}} \right]\left[ {\begin{array}{*{20}{c}}
1&0\\
-\frac{1}{f_2}&1
\end{array}} \right]\left[ {\begin{array}{*{20}{c}}
1&{{l_2}}\\
0&1
\end{array}} \right]\\
&=\left[ {\begin{array}{*{20}{c}}
\frac{{f_{1}}^2\,{f_4}^2-d\,\Delta_{l}\,f_3^2}{f_{1}\,f_{2}\,f_4\,f_3}&\frac{\Delta_{l}\,{f_{1}}^2\,{f_4}^2+d\,\Delta_{l}\,f_{1}\,f_3^2+\Delta_{l}\,{f_{2}}^2\,{f_3}^2-d\,l_{1}\,\Delta_{l}\,{f_3}^2}{f_{1}\,f_{2}\,f_4\,f_3}\\
-\frac{d\,f_3}{f_{1}\,f_{2}\,f_4}&\frac{f_3\,\left({f_{2}}^2-d\,\Delta_{l}\right)}{f_{1}\,f_{2}\,f_4}
\end{array}} \right] =  \left[ {\begin{array}{*{20}{c}}
{g_1}^*&{L}^*\\
-\frac{d\,f_3}{f_{1}\,f_{2}\,f_4}&{g_2}^*
\end{array}} \right]
\end{aligned}
\end{equation}
\end{figure*}

By calculating the single-pass transmission matrix of the SSLR and further obtaining the equivalent ${g_1}^*$ and ${g_2}^*$ parameters of the resonant cavity, the stability conditions of the resonant beam transfer system can be analyzed.
The single-pass transmission matrix ${\rm{M}}_s$ of an SSLR can be calculated using the ABCD matrix as shown in equation~(\ref{matrix1}).
Specifically, the symbol $l_1$ is the distance between L1 and M1; $l_2$ is the distance between L2 and M2; $f1$ and $f2$ is the focal length of lens L1 and L2 respectively; $\Delta_{l} = l_1 - f_1 = l_2 - f_2$ represents the assembly accuracy of the CER; $d$ represents the transmission distance from the pupil of the CER at the receiver to the pupil 2 of the telescope structure. 
When the single-pass transmission matrix of the SSLR satisfies the condition $0<{g_1}^*{g_2}^*<1$, the cavity is considered a stable cavity.
In this case, the stability condition of the SSLR can be represented by the range of transmission distance $d$ as
\begin{equation}
\label{range}
\left\{ \begin{array}{l}
d \in (0,\frac{{{{({n_1}{f_2})}^2}}}{{\Delta_l {n_2}^2}})~{\rm{if}}~{n_1} < {n_2}\\
d \in (0,\frac{{{f_2}^2}}{\Delta_l })~~~~~{\rm{if}}~{n_1} > {n_2}\\
d \in (0,2\frac{{{f_2}^2}}{\Delta_l })~~~~{\rm{if}}~{n_1} = {n_2}
\end{array} \right.,
\end{equation}
where $n_1 = \frac{f_1}{f_2}$.
According to the equation~(\ref{range}), without altering other parameters, there is a significant increase in the stable range of the resonant cavity when $n_1=n_2$.
Therefore, in designing an SSLR, we ensure that $n_1=n_2$ to maximize the transmission distance $d$ of the RBS.
At this point, the theoretical FoV of the AA-RBT $\theta_{\rm{A}}$ can be rewritten as
\begin{equation}
\label{theta1}
\theta_{\rm{A}} = {\rm{min}}\left[2{\rm{arctan}}(\frac{l_{t}}{f_4}), 2{\rm{arctan}}(\frac{{{l_{\rm{a}}}}}{{{f_2}}})\right].
\end{equation}
Therein, the latter variable represents the FoV of the CER at the receiver as in equation~(\ref{FoV01}).
Since the FoV of the system is affected by the FoV of the transmitter and receiver, the transmitter and receiver FoV are usually aligned to achieve the maximum system's FoV.
Therefore, we can consider that $\theta_{\rm{A}} =2{\rm{arctan}}(\frac{{{l_{\rm{a}}}}}{{{f_2}}})$.

\begin{figure}[!t]
\centering
\includegraphics[scale=1]{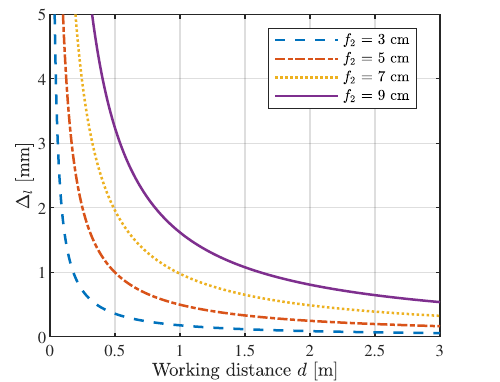}
\caption{Stability region (the area below the line) of $\Delta_l$ with different working distance $d$.}
\label{stab_range}
\end{figure}

When designing an SSLR with AA-RBT, both maximum transmission distance $d$ and the theoretical maximum FoV $\theta_{\rm{A}}$ are crucial factors to consider. 
Figure~\ref{stab_range} demonstrates calculated stability region of $\Delta_l$ versus transmission distance $d$ under different $f_2$. 
As the transmission distance $d$ increases, the stability region of $\Delta_l$ gradually decreases. 
At the same time, the smaller the value of $f_2$, the smaller the maximum transmission distance $d$ of the system under the same $\Delta_l$. 
The value of $\Delta_l$ is influenced by the precision of the installation of the AA-RBT, typically remaining below 0.1~mm in experimental setups.
Conversely, according to equation~(\ref{theta1}), the transmitter's FoV will increase as $f_2$ decreases.
Therefore, for a designed AA-RBT, there is a reciprocal constraint relationship between the system's maximum transmission distance and its FoV.

\subsection{Analysis on the RBS}

\subsubsection{Pumping threshold}
To determine whether a resonant beam can be output, it is essential to analyze whether the system transmission distance $d$ falls within a stable range and ascertain if the pump power surpasses the required threshold.
The threshold of the pump light power represents the minimum required for an SSLR to sustain oscillations and stably output the resonant beam at the receiver. 
The lower the pump light power threshold, the easier it is for the system to meet normal operating conditions.
The threshold $P_{\rm{th}}$ can be calculated based on the circulating power model~\cite{a181221.01}, which is depicted as

\begin{equation}
{P_{{\rm{th}}}} = \frac{{{A_{\rm{g}}}{I_{\rm{S}}}\left| {\ln \sqrt {{R_{\rm{r}}}{V_{\rm{a}}}^2{\eta _{\rm{d}}}} } \right|}}{{{\eta _{\rm{g}}}}},
\end{equation}
where $A_{\rm{g}}$ is the power stimulating area of the gain medium, which can be approximated as the area of the pump spot; $I_{\rm{S}}$ is the saturated light intensity of the gain medium; $R_{\rm{r}}$ is the reflectivity of the mirror M2; $V_{\rm{a}}$ is the transmission efficiency when the resonant beam passes through the optical component in the SSLR; $\eta _{\rm{d}}$ represents the round trip diffraction loss in the SSLR, and $\eta _{\rm{g}}$ is the excitation efficiency of the gain medium. 
Since an $\rm{Nd:YVO_4}$ crystal is used as the gain medium in the experiment, the corresponding crystal parameters $I_{\rm{s}}$, $\eta _{\rm{g}}$, and $\eta _{\rm{g}}$ can be easily obtained, as shown in Table~\ref{tab:my_label} of the Section~III.

The transmission efficiency $V_{\rm{a}}$ can be expressed as
\begin{equation}
    V_{\rm{a}} = R_{\rm{m}} \varGamma_1 \varGamma_2 \varGamma_3 \varGamma_4 R_{\rm{g}}
\end{equation}
where $R_{\rm{m}}$ is the reflectivity of mirror M1 and $R_{\rm{g}}$ is the reflectivity of the gain medium. $\varGamma_1$, $\varGamma_2$, $\varGamma_3$, and $\varGamma_4$ is the transmittance of lenses L1, L2, L3, L4, respectively. 
Since the surface of mirror M1 is coated with a high-reflective film, over 99.5\% of the light is reflected into the cavity, and only a minimal portion is absorbed or scattered by M1. 
All lenses in the SSLR are coated with anti-reflective film to reduce the loss of resonant beam. 
The losses caused by these factors are relatively stable as the receiver moves, so we assume that the transmission efficiency affected by these losses is approximately 88\%~\cite{9677003}.
However, the reflectivity of the gain medium changes significantly with the incident angle of the incoming light, as shown in Fig.~\ref{system01}(d). 
When the receiver moves radially, the incident angle changes noticeably, leading to variations in the reflectivity of the gain medium. 
The reduction in reflectivity significantly increases the threshold of pump optical power and reduces the end-to-end transmission efficiency of the resonant beam transfer system. 

\subsubsection{Output electrical power}

When the pump light power exceeds the pumping threshold, the resonant beam will be emitted from the output mirror at the receiver. 
A portion of the resonant beam is directed into the PV for energy harvesting through the PBS, while another portion is used for signal transmission.
Through the PV, the resonant beam can be converted into electrical energy for wireless charging.
The photocurrent generated when the resonant beam illuminates the PV is
\begin{equation}
    I_{\rm{P}} = \mu\zeta P_{\rm{out}},
\end{equation}
where $\mu$ is the split ratio of the PBS and $\zeta$ is the responsivity of PV.
The PV panel can be equivalent to a current source in parallel with a diode.
When a load with resistance $R_{\rm{L}}$ is connected to the PV panel, the output electrical power can be obtained as:
\begin{equation}
    \left\{ \begin{array}{l}
        {P_{\rm{PV}}} = {I_{\rm{out}}}^2{R_{\rm{L}}}\\
        {I_{\rm{out}}} = {I_{\rm{P}}} - {I_0}\left[ {{e^{\frac{{{I_{\rm{out}}}q({R_{\rm{L}}} + {R_{\rm{s}}})}}{{kTnn_{\rm{s}}}}}} - 1} \right] - \frac{{{I_{\rm{out}}}({R_{\rm{L}}} + {R_{\rm{s}}})}}{{{R_{\rm{sh}}}}}
        \end{array} \right.,
\end{equation}
where $q$ is the electron charge, $T$ is the temperature of the environment, $k$ is the Boltzmann constant, $I_0$ is the reverse saturation current, $n$ is the diode ideality factor, and $n_{\rm{s}}$ is the number of cells inside the PV panel. $R_{\rm{s}}$ and $R_{\rm{sh}}$ are the series resistance and the shunt resistance in the equivalent PV panel model, respectively.
Utilizing a maximum power point tracking device, the load resistance $R_{\rm{L}}$ of the PV can be adjusted to track the maximum power point of the PV, thus maximizing the electrical power output $P_{\rm{PV}}$ at the receiver.

\subsubsection{Bounds on Channel Capacity}

Since the resonant beam is the intracavity laser transmitted through free space, the resonant beam communication channel can be regarded as a free-space optical communication channel~\cite{lapidoth_capacity_2008}. 
Assuming the adoption of intensity modulation, the signal is modulated onto the resonant beam, and a PD is used to convert the received optical signal into a proportional electrical current at the receiver. 
Therefore, the electrical current signal $I_{\rm{A}}$ output from the PD can be depicted as
\begin{equation}
    I_{\rm{A}} = (1-\mu)\xi P_{\rm{out}},
\end{equation}
where $\xi$ is the responsivity of PD. 
Based on the free-space optical intensity channels theory~\cite{lapidoth_capacity_2008}, the lower bound on the capacity of the resonant beam communication channel can be obtained as 
\begin{equation}
    C(I_{\rm{A}}) = \frac{1}{2}{\rm{log}}\left( {1 + \frac{{{I_{\rm{A}}^2}e}}{{2\pi {\sigma ^2}}}} \right),
\end{equation}
where $\sigma ^2$ is the power of the noise.
Since the received signal is interfered with by a large number of independent sources under ambient light conditions, we model the noise as the sum of thermal noise $\sigma_{t}^2$ and shot noise $\sigma_{s}^2$:
\begin{equation}
    \left\{ \begin{array}{l}
        {\sigma ^2} = \sigma _t^2 + \sigma _s^2\\
        \sigma _t^2 = \frac{{4kTB}}{R}\\
        \sigma _s^2 = 2q({I_{\rm{A}}} + {I_{\rm{b}}})B
        \end{array} \right.
\end{equation}
where $B$ is the bandwidth of the signal, $I_b$ is the  background current, and $R$ is the load resistance.

\section{Experiments and Evaluations}
 
\subsection{Experimental Setup}

\begin{figure}[!t]
\centering
\includegraphics[scale=1]{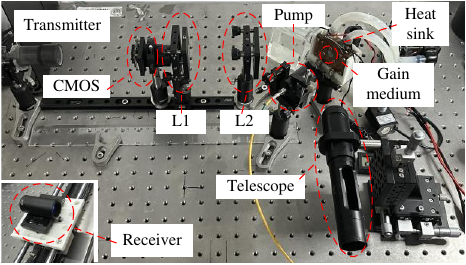}
\caption{Testbench setup for resonant beam transfer system.}
\label{fig}
\end{figure}

We construct an experimental platform to validate the proposed telescope-based resonant beam SLIPT system and measure the system output to analyze the system performance, as shown in Fig.~\ref{fig}.
In the transmitter of the resonant beam transfer system, a CER, composed of mirror M1 and lens L1, is placed on a linear guide.
The position of mirror M1 can be adjusted to precisely control the distance $\Delta_l$ between the M1 and L1.
The gain medium is located at the pupil of the CER and is arranged at a $45^\circ$ angle to the optical axis of the CER.
The collimator collimates the 808~nm pump light delivered via the fiber to form a spot with a beam radius of 0.8~mm.
The gain medium is a 1 mm thick $\rm{Nd:YVO_4}$ crystal slab connected to the semiconductor heat sink through a copper plate.
A telescope structure is mounted inside an optical sleeve and placed behind the gain medium.
At the bottom left of Fig.~\ref{fig} is the system's receiver, where lens L2 and mirror M2 are mounted inside a 5~cm long optical sleeve.
The receiver is placed on a linear guide slider parallel to the transmitter, allowing radial movement of the receiver and measurement of the optical power after mirror M2.
This setup enables precise measurement of the output optical power of the resonant beam transfer system at different radial angles of the receiver.
Table~\ref{tab:my_label} contains the additional parameters for the threshold of the pump light power calculation.

\begin{table}[ht]
    \centering
    \caption{Parameters in theoretical calculation~\cite{a181221.01}}
    \label{tab:my_label}
    \begin{tabular}{ccc}
    \toprule
    \textbf{Parameter} & \textbf{Symbol} & \textbf{Value} \\
    \midrule
    Saturation intensity & $I_{\rm{s}}$   & ${\rm{1.26}\times \rm{10^7~W/m^2}}$\\
    Reflectivity of the mirror M2 & $R_{\rm{r}}$ & 98.5\%\\
    Excitation efficiency & $\eta _{\rm{g}}$ & 72\%\\
    Area of the pump spot & $A_{\rm{g}}$ & ${\rm{8.04}\times \rm{10^6~m^2}}$\\
    Shunt resistance & $R_{\rm{sh}}$ & 53.82~$\Omega $\\
    Series resistance & $R_{\rm{s}}$ & 37~$m\Omega$\\
    PV's responsivity & $\zeta $ & 0.4~A/W\\
    PD's responsivity & $\xi $ & 0.6~A/W\\
    Reverse saturation current & $I_0$ & 0.32~$\mu$A\\
    background current & $I_{\rm{b}}$ & 5100~$\mu$A\\
    split ratio of the PBS & $\mu$ & 0.1\\
    Diode ideality factor & $n$ & 2.5\\
    Number of PV cell & $n_{\rm{s}}$ & 1\\
    \bottomrule
    \end{tabular}
\end{table}

\subsection{Performance Evaluation}

Based on theoretical analysis and experimental measurements, we obtain the output optical power and theoretical communication efficiency of the telescope-based RBS system. 
We also analyze the impact of the gain medium reflectivity on system performance and determine the transmitter's FoV of the system through the experiment.

\subsubsection{Transmission performance of the telescope-based RBS}

\begin{figure*}[htbp]	
    \centering
	\subfigure[Output optical power $P_{\rm{out}}$] %第一张子图
	{
		\begin{minipage}[t]{0.3\textwidth}
			\centering          %子图居中
			\includegraphics[scale=0.7]{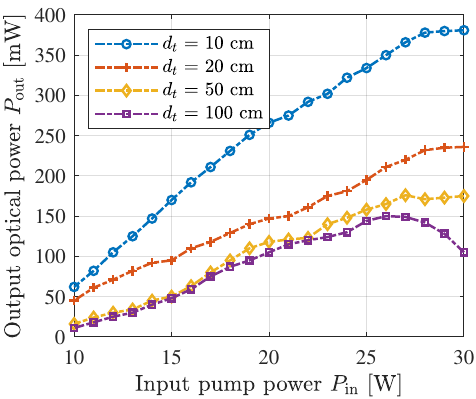}   
		\end{minipage}
	}
    \hspace{0mm} % 在两个minipage之间添加一些水平空间
	\subfigure[Output electrical power $P_{\rm{PV}}$] %第二张子图
	{
		\begin{minipage}[t]{0.3\textwidth}
			\centering      %子图居中
			\includegraphics[scale=0.7]{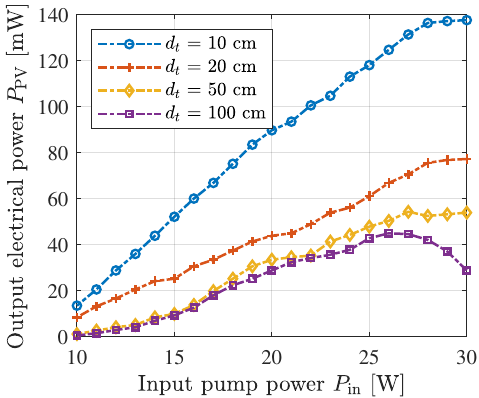}   
		\end{minipage}
	}
    \hspace{0mm} % 在两个minipage之间添加一些水平空间
	\subfigure[Spectral efficiency $C$] %第二张子图
	{
		\begin{minipage}[t]{0.3\textwidth}
			\centering      %子图居中
			\includegraphics[scale=0.7]{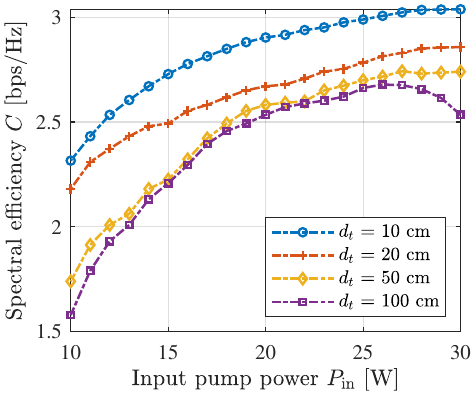}   
		\end{minipage}
	}
	\caption{Output optical power, output electrical power, and spectral efficiency as a function of input pump power at different transmission distances $d_{\rm{t}}$. } %  %大图名称
	\label{p_pump_vs_p_out}  %图片引用标记
\end{figure*}

Figure~\ref{p_pump_vs_p_out}(a) demonstrates the curve of the output resonant optical power of the resonant beam $P_{\rm{out}}$ at receiver with the change in input pump light power $P_{\rm{in}}$ at different transmission distances. 
When the transmission distance $d = 10$~cm, the output optical power increases with the increase of the input pump light power. 
Since the focal lengths of lenses L2 and L4 are 5~cm, the optical pupils of the transmitter and receiver coincide. 
In this case, the thermal lens effect caused by the increased power has a lower impact on the output of the resonant beam. 
The output optical power is higher than that under other transmission distances and continues to increase as the pump light power increases.
When the transmission distance $d_{\rm{t}}$ is set at 20~cm, 50~cm, and 100~cm, respectively, the output optical power $P_{\rm{out}}$ initially increases and then slowly decreases with the increase in pump light power. 
This is because as the pump light power increases, the impact of the thermal lens effect of the gain medium becomes more significant, causing a continuous reduction in the focal length of the gain medium. 
This results in an increase in the size of the resonant beam spot inside the cavity, thereby increasing the transmission loss of the resonant beam in the cavity.

Based on the resonant beam optical power measured at the receiver from the experiments, we have calculated the theoretical transmission electrical power and channel capacity of the resonant beam SLIPT system, as is depicted in Figs.~\ref{p_pump_vs_p_out}(b) and (c). 
The results indicate that, with the telescope-based RBS, it's possible to provide a continuous power supply of up to 45~mW to the receiving device from a distance of 1~m. 
At the same time, real-time communication can be achieved through the system, with a spectral efficiency of up to 2.7~bps/Hz.

\subsubsection{Impact of the gain medium reflectivity on pumping threshold}

\begin{figure}[!t]
\centering
\includegraphics[scale=1]{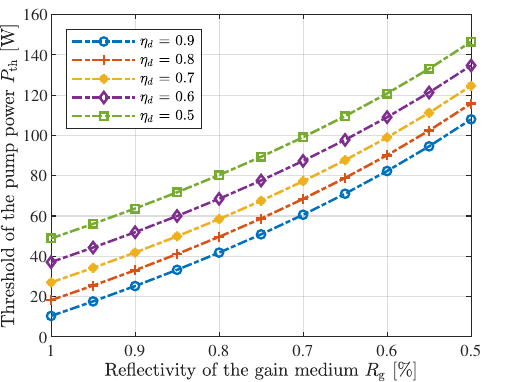}
\caption{Threshold of the pump power $vs.$ reflectivity of the gain medium under different round trip diffraction loss.}
\label{p_pump_vs_R_gain}
\end{figure}

According to the circulating power model, the relationship between the threshold of the pump light power $P_{\rm{th}}$ and the reflectivity of the gain medium $R_{\rm{g}}$ can be calculated as demonstrated in Fig.~\ref{p_pump_vs_R_gain}.
As the round trip loss factor $\eta_{\rm{d}}$ gradually decreases from 0.9 to 0.5, the threshold power continuously increases.
With the ongoing reduction in the reflectivity of the gain medium, the threshold power keeps rising. 
When the incident angle of the resonant beam on the gain medium increases from $0^\circ$ to $15^\circ$, the reflectivity of the gain medium will decrease about 15\% as shown in Fig.~\ref{system01}(d).
Correspondingly, a 15\% decrease in reflectivity approximately results in a 20~W increase in the threshold of the pump light power.
This poses higher requirements for the pump light power of the resonant beam transfer system and significantly reduces the transmission efficiency of the system.
More importantly, a greater pump light power can lead to a more severe thermal lens effect, further reducing the transmission efficiency of the SSLR and shrinking the stable region of the SSLR.
As a result, the radial and axial movement distance of the receiver will be limited.

\begin{figure}[!t]
\centering
\includegraphics[scale=1]{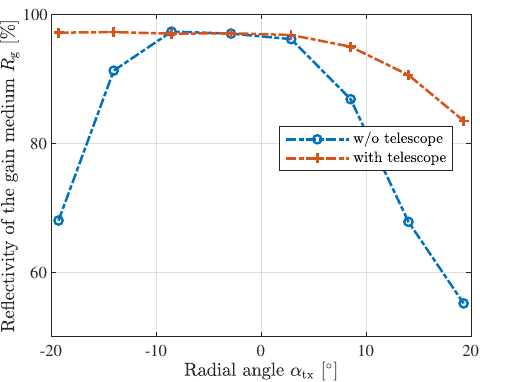}
\caption{Reflectivity of the gain medium $vs.$ radial angle of the receiver with and without a telescope structure.}
\label{refl_vs_angle}
\end{figure}

The reflectivity of the gain medium with and without a telescope structure in the variation of radial angle $\alpha_{\rm{tx}}$ is depicted in Fig.~\ref{refl_vs_angle}.
On the experimental testbed, the angle between the gain medium and the optical axis of the CER at the transmitter is $45^\circ$. 
Therefore, when the radial angle is $0^\circ$, the receiver is perfectly aligned with the transmitter, and the angle between the intra-cavity resonant beam and the gain medium is $45^\circ$. 
As the receiver moves radially relative to the transmitter, the radial angle changes continuously, causing a change in the angle between the intra-cavity resonant beam and the gain medium, further affecting the reflectivity of the gain medium. 
Without a telescope structure in the cavity, the gain medium can only maintain a reflectivity above 90\% when the radial angle $\alpha_{\rm{tx}}\in(-15,5)$ degrees. 
Once the radial angle $\alpha_{\rm{tx}}$ continues to increase, the reflectivity quickly drops below 70\%, resulting in a large amount of losses in the resonant cavity.
Eventually, the gain medium will not be able to provide enough gain to maintain the oscillation in the resonant cavity, and the resonant beam will not be generated.
Therefore, when the radial angle increases beyond $10^\circ$, the FoV of the system is limited by the reflectivity of the gain medium, making it challenging to enhance further.
However, by adding a telescope structure to the transmitter, which includes lenses L3 and L4 with focal lengths of 100~mm and 50~mm, the angle of the resonant beam entering the transmitter is reduced by half before it reaches the gain medium.
In this case, the gain medium can maintain a reflectivity above 90\% for radial angle $\alpha_{\rm{tx}}\in(-20,15)$ degrees.

\begin{figure*}[htbp]	
    \centering
	\subfigure[Movement of the resonant beam spot on the CMOS (without telescope).] %第一张子图
	{
		\begin{minipage}[t]{0.9\textwidth}
			\centering          %子图居中
			\includegraphics[scale=0.9]{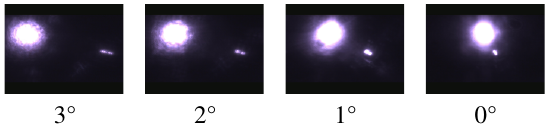}   
		\end{minipage}
	}
	\subfigure[Movement of the resonant beam spot on the CMOS (with telescope).] %第二张子图
	{
		\begin{minipage}[t]{1\textwidth}
			\centering      %子图居中
			\includegraphics[scale=0.9]{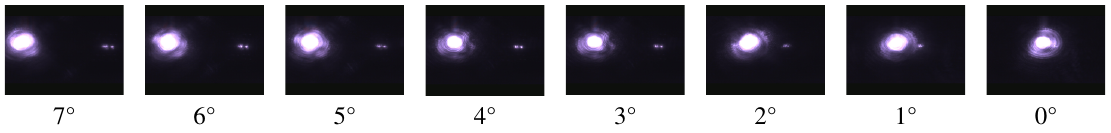}   
		\end{minipage}
	}
	\caption{As the radial angle $\alpha_{\rm{tx}}$ of the receiver changes, the movement of the resonant beam spot on the CMOS.} %大图名称
	\label{spot}  %图片引用标记
\end{figure*}

To verify whether incorporating a telescope into the RBS transmitter could effectively reduce the angle at which the resonant beam impinges on the gain medium, we collected images of the resonant beam spot on the receiver end using a CMOS sensor~\cite{xu_binocular_2023,liu_passive_2023}, as illustrated in Fig.~\ref{spot}. 
As the receiver moves, its radial angle relative to the transmitter's optical axis changes, causing the location of the resonant beam spot on the CMOS to shift. 
At small radial angles, the movement of the light spot on the CMOS can be considered proportional to the change in the incidence angle of the resonant beam on the gain medium. 
When the radial angle is 0$^\circ$, the resonant beam spot is located at the center of the CMOS. 
For an RBS transmitter without a telescope, as the radial angle gradually increases to 3$^\circ$, the light spot moves towards the edge of the CMOS. 
In contrast, for an RBS transmitter equipped with a telescope, the same distance movement of the light spot on the CMOS occurs only when the radial angle increases to 6$^\circ$. 
This indicates that the telescope in the RBS transmitter halved the range of change in the angle between the resonant beam and the gain medium.

\subsubsection{Transmitter's FoV of the telescope-based RBS}

\begin{figure}[!t]
\centering
\includegraphics[scale=1]{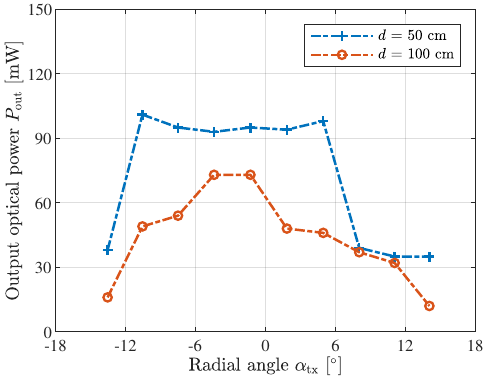}
\caption{Output optical power $vs.$ radial angle of the receiver under different transmission distances. The optical power of the pump light $P_{\rm{in}}=20$~W }
\label{RadialAngle_vs_P_out}
\end{figure}

The relationship between the output optical power of resonant beam at receiver $P_{\rm{out}}$ and the radial angle $\alpha_{\rm{tx}}$ at different transmission distances is measured, as shown in Fig.~\ref{RadialAngle_vs_P_out}. 
Since the focal lengths of lenses L2 and L4 are 50~mm, with a diameter of 25.4~mm, the theoretical limit of the system's FoV can be calculated as $28.5^\circ$ by equation~(1). 
It is observed that at transmission distances of 50~cm and 100~cm, the system's maximum FoV is essentially consistent.
The output optical power $P_{\rm{out}}$ remains stable when the radial angle changes from $-14^\circ$ to $14^\circ$, reaching a FoV of $28^\circ$, which is near the theoretical limit.
Additionally, as the transmission distance increases, the transmission loss in the resonant cavity also increases, leading to lower output optical power at the same radial angle over 100~cm transmission distance. 
The incorporation of a telescope structure in the system results in a more gradual change in the reflectivity of the gain medium during the radial movement of the receiver. 
Therefore, the fluctuation in the output optical power at the receiver is not significant during the change in radial angle. 
When the radial angle approaches the theoretical maximum FoV, the output optical power will quickly drop to 0~w.

\begin{table*}[ht]
    \centering
    \caption{System performance comparison}
    \label{tab:my_label2}
    \begin{tabular}{ccccccc}
    \toprule
    \textbf{Ref.}  & \textbf{Retro-reflector} &\textbf{Gain medium} &\textbf{Working distance (m)}& \textbf{Transmitter FoV($^\circ$)} & \textbf{Year}\\
    \midrule
     \cite{Lim:19}& Corner cube  &Semiconductor chip& 1~m & \makecell[c]{$6.6^\circ$ \\(to 0 output)} & 2019\\
     \cite{Li:18}& CER  &${\rm{Nd:YVO_4}}$ disk& 0.15~m & \makecell[c]{$\pm 8.3^\circ$ \\(to 0 output)} & 2021\\
     \cite{9677003}& CER  &${\rm{Nd:YVO_4}}$ disk& 2~m & \makecell[c]{$\pm 5.1^\circ$ \\(to 0 output)} & 2022\\
     \cite{Javed:22}& CER  &Optical pumped VECSEL& 2~m & \makecell[c]{$\pm 0.47^\circ$ \\(to 0 output)} & 2022\\
     \cite{sheng_adaptive_2023} & CER  & Bulk ${\rm{Nd:YVO_4}}$& 5~m & \makecell[c]{$4.6^\circ$ \\(half maximum)} & 2023\\
    This work & CER & ${\rm{Nd:YVO_4}}$ disk & 1~m & \makecell[c]{$28^\circ$ \\(to 0 output)}  & 2024\\
    \bottomrule
    \end{tabular}
\end{table*}

To demonstrate the advantages of our proposed system, we have delineated in Table~\ref{tab:my_label2} the principal performance characteristics of recent self-aligning resonant beam wireless transmission schemes that utilize SSLR. 
Systems employing Bulk crystals as the gain medium have historically faced challenges in achieving an extensive transmitter FoV. 
In~\cite{sheng_adaptive_2023}, the implementation of CERs to compensate for spherical aberration and field curvature effectively augmented the receiver FoV. 
However, the transmitter FoV, which is of greater significance, remained considerably limited. 
Systems utilizing disk gain media have more readily achieved a broader transmitter FoV, yet they still fall considerably short of the theoretical maximum FoV. 
In contrast, our study has accomplished a 28$^\circ$ FoV at the transmitting end over meter-level transmission distances. 
This achievement represents at least a doubling of the transmitter FoV relative to prior research, nearing the theoretical limit of FoV.

\section{Discussions}

\subsection{Trade-off between System's FoV and Efficiency}
Incorporating a telescopic structure into the transmitter of a resonant beam transmission system effectively enhances the transmitter FoV. 
Additionally, as the focal length ratio of the lenses in the telescopic structure increases, the incident angle of the resonant beam on the gain medium gradually decreases. 
Consequently, a larger focal length ratio leads to reduced losses due to the reflectivity of the gain medium. 
However, on the other hand, as the focal length ratio increases, the divergence angle of the resonant beam in the SSLR increases, resulting in an increase in diffraction loss during spatial transmission, thereby increasing the overall transmission loss of the system.
Therefore, when designing a resonant beam transmission system, it's crucial to balance the losses at the gain medium and the losses during spatial transmission. 
By adjusting the focal length ratio of the lenses, the system's FoV and transmission efficiency can be optimized at an appropriate transmission distance. 
This delicate balance is key to maximizing the performance of the system, ensuring that improvements in FoV do not come at the massive expense of transmission efficiency.

\subsection{Further Improvements Scheme}
Utilizing a solid-state gain medium in the construction of SSLRs often leads to unavoidable optical losses in the variation of the incident angle due to the structural characteristics of the solid-state laser.
Employing vertical external cavity surface emitting lasers (VECSEL) can effectively reduce these additional losses due to variations in the incident angle while substantially decreasing the system's volume, thereby enhancing its integrability.
Additionally, replacing optical pumping with electrical pumping in VECSELs can increase energy conversion efficiency, thus improving the end-to-end efficiency of the system. 
VECSELs are capable of maintaining stable performance over extended operational periods and allow for precise output power control by adjusting the electric current, facilitating automation and accurate control. 
Therefore, replacing solid-state lasers with VECSELs in the construction of resonant beam transmission systems represents a promising direction for future enhancements.

\section{Conclusions}

In this paper, we conduct a detailed analysis of the key factors affecting the FoV in resonant beam transmission systems and propose a system design that enhances the FoV of the transmitter in resonant beam wireless transmission systems. 
By incorporating a telescopic structure into the transmitter of the resonant beam transmission system, we amplify the angle of the resonant beam incident on the gain medium and direct it toward the receiver. 
This design increases the reflectivity of the resonant beam at the gain medium, thereby improving the system's transmission efficiency at large viewing angles and consequently enhancing the transmitter's FoV. 
Experimental results demonstrate that our designed resonant beam transmission system increases the system's transmitter FoV to 28$^\circ$. 
Our work paves the way for applications of resonant light transmission systems in resonant beam charging systems and resonant beam communication systems.

\ifCLASSOPTIONcaptionsoff
  \newpage
 \fi

\bibliographystyle{IEEEtran}
\small

\bibliography{mybib}

\end{document}